\documentstyle[12pt]{article}
\def\be{\begin{equation}}
\def\ee{\end{equation}}
\def\bea{\begin{eqnarray}}
\def\eea{\end{eqnarray}}

\begin{document}

\begin{center}
{\Large{\bf Moving Branes with Background Massless and Tachyon
Fields in the Compact Spacetime}}

\vskip .5cm {\large Zahra Rezaei and Davoud Kamani}
\vskip .1cm
{\it Physics Department, Amirkabir University of Technology
(Tehran Polytechnic)
\\  P.O.Box: 15875-4413, Tehran, Iran}\\
{\sl e-mails: z.rezaei , kamani@aut.ac.ir}\\
\end{center}

\begin{abstract}

In this article we shall obtain the boundary state associated with
a moving $Dp$-brane in the presence of the Kalb-Ramond field
$B_{\mu\nu}$, an internal $U(1)$ gauge field $A_{\alpha}$ and a
tachyon field, in the compact spacetime. According to this state,
properties of the brane and a closed string, with
mixed boundary conditions emitted from it, will be obtained. Using
this boundary state we calculate the interaction amplitude of two
moving $Dp_{1}$ and $Dp_{2}$-branes with above background fields
in a partially compact spacetime. They are parallel or perpendicular to
each other. Properties of the interaction amplitude will be analyzed and
contribution of the massless states to the interaction will be extracted.

\end{abstract}



\vskip .5cm
\newpage

\section{Introduction}

Strings are not the only objects of string theory. Since 1995 it has
been cleared \cite{1} that the theory includes extended objects
which carry charges related to special antisymmetric fields which
string theory describes but can not be source of them. These are
D-branes which are found to be important in nonperturbative string
theories because in strong coupling they become arbitrarily light
(lighter than string itself) and dominate the theory in low energies
\cite{2}.

One of the interesting subjects about the D-branes is interaction
between them which is obtainable through two different but
equivalent procedures: one loop diagram of open string and tree
level diagram of closed string \cite{3}. Since two D-branes
interaction can be described by exchanging of closed strings, here
we restrict ourselves to the second approach. The state which
describes the closed string production from vacuum is called
boundary state. Boundary state formalism is a powerful method for
studying branes properties and their interactions.

Among achievements in the subject of boundary state formalism is
studying the interaction of mixed branes (branes with both Neumann
and Dirichlet boundary conditions), moving and angled branes in the
presence of background fields such as U(1) gauge field \cite{4} and
antisymmetric field $B_{\mu\nu}$ \cite{5, 6, 7, 8, 9,10,11,12,13}.
Tachyon field also has been added as a background field in some
studies \cite{14,15,16}.

Since D-branes are not static objects, studying their dynamics is
essential to interpret them as physical objects in string theory.
Considering velocity for a D-brane which is equal to taking into
account the scalar fields from the worldsheet point of view
\cite{17}, as well as gauge field on the D-brane worldvolume is very
instructive to study D-branes dynamics. Besides, progresses in
studying open string tachyon field which began mainly by the Sen's
works \cite{15,18,19} show that this tachyon field plays an
important role on improving our knowledge about D-branes, their
instability or stability features, true vacuum of tachyonic string
theories and etc \cite{18}.

The above facts motivated us to study a system of two moving
D$p_{1}$ and D$p_{2}$-branes in the presence of the following
background: tachyon field, Kalb-Ramond field $B_{\mu\nu}$, $U(1)$
gauge fields which live in the worldvolumes of the branes, and a
partially compacted spacetime on tori. The branes dimensions $p_{1}$
and $p_{2}$ are arbitrary. The relative configurations of the branes
are parallel and/or perpendicular. Without fixing the position of
the branes, we study both configurations simultaneously. We
calculate the boundary state, corresponding to the branes, and then
obtain the interaction amplitude between them through exchange of
closed strings. While the spacetime is allowed to have some compact
directions we observe that presence of the tachyon field has some
effects on wrapping of the closed string around these directions. In
addition, the tachyon field also affects the interaction amplitude
of the branes. For example, the behavior of the amplitude for large
distance branes has a major deviation from what is expected in the
conventional case which will be interpreted.
\section{The boundary state}

We begin with a special sigma-model for the string.
This sigma-model action contains the
antisymmetric field $B_{\mu\nu}$, tachyon fields, two $U(1)$
gauge fields which live on the worldvolume of the branes and two
velocity terms corresponding to the motion of the branes
\bea
S=&~& -\frac{1}{4\pi\alpha'} {\int}_{\Sigma} d^{2}\sigma
(\sqrt{-g}g^{ab}G_{\mu\nu}\partial_{a}X^{\mu}\partial_{b}X^{\nu}
+\varepsilon^{ab}
B_{\mu\nu}\partial_{a}X^{\mu}\partial_{b}X^{\nu})
\nonumber\\
&~& -\frac{1}{2\pi\alpha'} {\int}_{(\partial\Sigma)_{1}} d\sigma
\bigg{(} A^{(1)}_{\alpha_{1}}
\partial_{\sigma}X^{\alpha_{1}}+V_{1}^{i_{1}}X^{0}\partial_{\tau}
X^{i_{1}}+(T^{(1)}
+\frac{1}{2}U^{(1)}_{\mu\nu}X^{\mu}X^{\nu})\bigg{)}
\nonumber\\
&~& +\frac{1}{2\pi\alpha'} {\int}_{(\partial\Sigma)_{2}} d\sigma
\bigg{(}A^{(2)}_{\alpha_{2}}
\partial_{\sigma}X^{\alpha_{2}}+V_{2}^{i_{2}}X^{0}\partial_{\tau}
X^{i_{2}}
+(T^{(2)}+\frac{1}{2}U^{(2)}_{\mu\nu}X^{\mu}X^{\nu})\bigg{)},
\eea
where $\Sigma$ is the worldsheet of the closed string, exchanged
between the branes. The boundaries of this worldsheet, i.e.
${(\partial\Sigma)}_{1}$ and ${(\partial\Sigma)}_{2}$, are at
$\tau=0$ and $\tau=\tau_{0}$, respectively. The $U(1)$ gauge field
$A^{(2)}_{\alpha_{2}}$ lives in the D$p_{2}$-brane,
and $V_{2}^{i_{2}}$ is its velocity component along
$X^{i_{2}}$ direction. The set $\{X^{\alpha_{2}}\}$ specifies the
directions along the D$p_{2}$-brane worldvolume and
$\{X^{i_{2}}\}$ shows the directions perpendicular to it.
Similar variables with the index ``1'' refer to the D$p_{1}$-brane.

Here we take the background fields $G_{\mu\nu}$ and $B_{\mu\nu}$ to
be constant and the profile of the tachyon field is defined as
$T^2=T_{0}+\frac{1}{2}U_{\mu\nu}X^{\mu}X^{\nu}$ with constant
$T_{0}$ and constant symmetric matrix $U_{\mu\nu}$. The advantage of
this profile is that the theory will be Gaussian and therefore is
exactly solvable \cite{16}. Vanishing the variation of this action
with respect to $X^{\mu}(\sigma,\tau)$ gives the equation of motion
of $X^{\mu}(\sigma,\tau)$ and boundary state equations of the
emitted (absorbed) closed string from (by) the brane. For
simplicity, we remove the indices ``1'' and ``2'' of the variables
which refer to the D$p_{1}$ and D$p_{2}$-branes. In the interaction
of these branes again we shall restore these indices. Therefore, we
receive the mixed boundary state equations (i.e. a combination of
Dirichlet and Neumann boundary conditions) at $\tau=0$,
\begin{equation}
[\partial_{\tau} (X^{0}-V^{i}X^{i}) + {\cal
{F}}^{0}_{\;\;\;\alpha}\partial_{\sigma}X^{\alpha}
-U^{0}_{\;\;\;\nu}X^{\nu}]_{\tau=\tau_{0}} |B_{x}, \tau=0\rangle
=0,
\end{equation}
\begin{equation}
[\partial_{\tau} X^{\bar{\alpha}} + {\cal
{F}}^{\bar{\alpha}}_{\;\;\; \beta}\partial_{\sigma}X^{\beta}
-U^{\bar{\alpha}}_{\;\;\;\nu}X^{\nu}]_{\tau=0} |B_{x},
\tau=0 \rangle =0,
\end{equation}
\begin{equation}
[X^{i}-V^{i}X^{0}-y^{i}]_{\tau=0}|B_{x}, \tau=0\rangle =0,
\end{equation}
where $\bar{\alpha}$ refers to the spatial directions of the brane
(i.e. $\bar{\alpha}\neq0$), $\{y^i\}$ denotes the initial transverse
coordinates of the brane, and $\cal {F}$ is total field strength
\begin{equation}
{\cal{F}}_{\alpha\beta}=\partial_{\alpha}A_{\beta}-\partial_{\beta}
A_{\alpha}-B_{\alpha\beta}. \label {5}
\end{equation}
The first two terms define the field strength of $A_\alpha$,
which is assumed to be constant. For simplification, we also
assumed that the mixed elements of the
Kalb-Ramond field to be zero, i.e., $B^{\alpha}_{\;\;i}=0$.

To solve these equations we use the general solution of the closed
string equation of motion
\begin{equation}
X^{\mu}(\sigma,\tau)=x^{\mu}+2\alpha^{'}p^{\mu}\tau+2L^{\mu}\sigma
+\frac{i}{2}
\sqrt{2\alpha^{'}}\sum_{m\neq0}\frac{1}{m}(\alpha^{\mu}_{m}
e^{-2im(\tau-\sigma)}
+\tilde{\alpha}^{\mu}_{m}e^{-2im(\tau+\sigma)}).
\end{equation}
In this relation $L^{\mu}$ is zero for non-compact directions and
$L^{\mu}=N^{\mu}R^{\mu}$ for compact directions where $N^{\mu}$ is
the winding number of the closed string and $R^{\mu}$ is the
radius of compactification of the compact direction $X^{\mu}$. Therefore,
the boundary state Eqs. (2)-(4) can be written in
terms of the oscillators
\bea
&~&[(\alpha^{0}_{m}-V^{i}\alpha^{i}_{m}- {{\cal
{F}}^{0}}_{\alpha}\alpha^{\alpha}_{m}-\frac{i}{2m}U^{0}_{\;\;\;\nu}
\alpha^{\nu}_{m})
\nonumber\\
&~&+(\tilde{\alpha}^{0}_{-m}-V^{i}\tilde{\alpha}^{i}_{-m} +{{\cal
{F}}^{0}}_{\alpha}\tilde{\alpha}^{\alpha}_{-m}+\frac{i}{2m}
U^{0}_{\;\;\;\nu}\tilde{\alpha}^{\nu}_{-m})] |B_{x},
\tau=0\rangle =0,
\eea
\bea
&~&[(\alpha^{\bar{\alpha}}_{m}-
{{\cal {F}}^{\bar{\alpha}}}_{\beta}\alpha^{\beta}_{m}-\frac{i}{2m}
U^{\bar{\alpha}}_{\;\;\;\nu}\alpha^{\nu}_{m})
\nonumber\\
&~& +(\tilde{\alpha}^{\bar{\alpha}}_{-m}+
{{\cal {F}}^{\bar{\alpha}}}_{ \beta}\tilde{\alpha}^{\beta}_{-m}+
\frac{i}{2m}U^{\bar{\alpha}}_{\;\;\;\nu}
\tilde{\alpha}^{\nu}_{-m})] |B_{x}, \tau=0\rangle =0,
\eea
\bea
&~&[(\alpha^{i}_{m}-V^{i}\alpha^{0}_{m})
-({\tilde{\alpha}}^{i}_{-m}-V^{i}{\tilde{\alpha}}^{0}_{-m})]|B_{x},
\tau=0\rangle =0,
\eea
and zero modes
\begin{equation}
[2\alpha^{'}(p^{0}-V^{i}p^{i})+2{{\cal
{F}}^{0}}_{\alpha}L^{\alpha}-U^{0}_{\;\;\;\nu}x^{\nu})_{\rm op}]|B_{x},
\tau=0\rangle =0,
\end{equation}
\begin{equation}
U^{0}_{\;\;\;\nu}L_{\rm op}^{\nu}|B_{x}, \tau=0\rangle =0,
\end{equation}
\begin{equation}
(2\alpha^{'}p^{\bar{\alpha}}+2{{\cal
{F}}^{\bar{\alpha}}}_{\beta}L^{\beta}
-U^{\bar{\alpha}}_{\;\;\;\nu}x^{\nu})_{\rm op}|B_{x}, \tau=0\rangle =0,
\end{equation}
\begin{equation}
U^{\bar{\alpha}}_{\;\;\;\nu}L_{\rm op}^{\nu}|B_{x}, \tau=0\rangle =0,
\end{equation}
\begin{equation}
(x^{i}-V^{i}x^{0}-y^{i})_{\rm op}|B_{x}, \tau=0\rangle =0,
\end{equation}
\begin{equation}
L_{\rm op}^{i}|B_{x}, \tau=0\rangle =0.
\end{equation}
\begin{equation}
(p^{i}-V^{i}p^{0})_{\rm op}|B_{x}, \tau=0\rangle =0.
\end{equation}
where we assumed that the time direction is non-compact, i.e.
$L^{0}=0$. The index ``op'' means that the variables are operator.

From now on we put a restriction on the velocities and consider
both D$p_{1}$ and D$p_{2}$-branes to move along the common
direction $X^{i_{0}}$, which is perpendicular to both of them. Thus,
the Eqs. (10), (12) and (16), lead to
\bea
&~&p^{0}=\frac{\gamma^{2}}{\alpha'}(\frac{1}{2}
U^{0}_{\;\;\;\mu}x^{\mu}-{{\cal
{F}}^{0}}_{\bar{\beta}}L^{\bar{\beta}}),
\nonumber\\
&~& p^{\bar{\alpha}}=\frac{1}{\alpha'}(\frac{1}{2}
U^{\bar{\alpha}}_{\;\;\;\mu}x^{\mu}-{{\cal
{F}}^{\bar{\alpha}}}_{\bar{\beta}}L^{\bar{\beta}}),
\nonumber\\
&~&p^{i_{0}}=\frac{V\gamma^{2}}{\alpha'}(\frac{1}{2}
U^{0}_{\;\;\;\mu}x^{\mu}-{{\cal
{F}}^{0}}_{\bar{\beta}}L^{\bar{\beta}}),
\eea
where $\gamma=1/\sqrt{1-V^{2}}$. For the compact direction $X^\mu$
we also have $p^\mu=M^\mu/R^\mu$ where $M^\mu$ is the momentum number
of the closed string. These equations imply that the nonzero momentum
components (momentum numbers) of closed string depend on its winding
numbers around the wrapped directions of the brane, and its center of
mass position. The former is due to the massless background fields,
while the latter is the effect of the tachyon field.

Combining the Eqs. (11), (13) and (15) leads to
\bea
U^{\alpha}_{\;\;\;\bar{\beta}}L_{\rm op}^{\bar{\beta}}|B_{x},
\tau=0\rangle =0.
\eea
If the $p\times p$ square sub-matrix
$U^{\bar{\alpha}}_{\;\;\;\bar{\beta}}$ is invertible we obtain
\bea L_{\rm op}^{\bar{\alpha}}|B_{x}, \tau=0\rangle =0.
\eea
Thus, the background tachyon field prevents the wrapping
of the closed string around the compact directions of the brane.
The Eqs. (15) and (19) imply that the closed string
can not wrap around any compact direction of the spacetime.
Putting aside the invertibility of the sub-matrix
$U^{\bar{\alpha}}_{\;\;\;\bar{\beta}}$, the closed string can wind
around the compact directions of the brane. However, we assume
$U^{\bar{\alpha}}_{\;\;\;\bar{\beta}}$ to be invertible.

Now we solve the boundary state equations to obtain the boundary
state. By using the coherent state method \cite{20}, the Eqs.
(7)-(9) give the oscillating part of the boundary state as in the
following \bea |B_{\rm osc},\tau=0\rangle=\prod^{\infty}_{n=1}[\det
M_{(n)}]^{-1}\exp\bigg {[}-\sum_{m=1}^{\infty}\bigg
{(}\frac{1}{m}\alpha^{\mu}_{-m}{\cal{S}}_{(m)\mu\nu}
\widetilde{\alpha}^{\nu}_{-m}\bigg {)}\bigg{]}|0\rangle, \eea where
\bea \left\{ \begin{array}{rcl} &~&
{\cal{S}}_{(m)}=S_{(m)}+((S_{(-m)})^{-1})^{T},\\
&~& S_{(m)}=M^{-1}_{(m)}N_{(m)}.
\end{array}\right.
\eea
Since $S_{(m)}$ is mode dependent and generally is not
orthogonal, the matrix $((S_{(-m)})^{-1})^{T}$ is appeared here.
The matrices $M_{(m)}$ and $N_{(m)}$, which depend on ${\cal
{F}}$, $V$, $B$ and $U$, are defined by
\bea
M^{\mu}_{(m)\nu}=\Omega^{\mu}_{\;\;\nu}-\frac{i}{2m}
U^{\alpha}_{\;\;\nu}\delta^{\mu}_{\;\;\alpha}
\eea
where
\bea
\left\{
\begin{array}{rcl} &~&\Omega^{0}_{\;\;\mu}=\delta^{0}_{\;\;\mu}
-V{\delta^{i_{0}}}_{ \mu}-{{\cal
{F}}^{0}}_{\alpha}{\delta^{\alpha}}_{
\mu}, \\
&~&
\Omega^{\bar{\alpha}}_{\;\;\mu}=\delta^{\bar{\alpha}}_{\;\;\mu}-{{\cal
{F}}^{\bar{\alpha}}}_{\beta}{\delta^{\beta}}_{
\mu}, \\
&~& \Omega^{i}_{\;\;\mu}=\delta^{i}_{\;\;\mu}
-V{\delta^{i}}_{\;i_{0}}{\delta^{0}}_{ \mu},
\end{array}\right.
\eea
and
\bea
\left\{ \begin{array}{rcl}
&~&N^{0}_{(m)\mu}={\delta^{0}}_{ \mu}-V{\delta^{i_{0}}}_{\;\mu}+
{{\cal {F}}^{0}}_{\alpha}{\delta^{\alpha}}_{
\mu}+\frac{i}{2m}U^{0}_{\;\;\;\mu}, \\
&~& N^{\bar{\alpha}}_{(m)\mu}={\delta^{\bar{\alpha}}}_{
\mu}+{{\cal {F}}^{\bar{\alpha}}}_{\beta}{\delta^{\beta}}_{
\mu}+\frac{i}{2m}U^{\bar{\alpha}}_{\;\;\;\mu}, \\
&~& N^{i}_{(m)\mu}=-{\delta^{i}}_{
\mu}+V{\delta^{i}}_{\;i_{0}}{\delta^{0}}_{\;\mu},
\end{array}\right.
\eea
where $V^{i_{0}}\equiv V$. The infinite factor in (20) can be
regularized as
\bea
\prod^{\infty}_{n=1}[\det M_{(n)}]^{-1}=\sqrt{\det
\Omega}\;\det\Gamma \bigg{(}\frac{U}{2i\Omega}+1\bigg{)}.
\eea
It is seen that if $U=0$ then (25) will be the familiar DBI Lagrangian.

Solving the zero mode equations (10)-(16) by considering $x^\mu$ and
$p^\mu$ as quantum mechanical operators and using their commutation
relations, we receive the state
\bea
{|B_{x},\tau=0\rangle}^{(0)}=&~&\frac{T_{p}}{2}
{\int^{\infty}_{-\infty}}\prod_{\mu}dp^{\mu}\;\;\bigg{\{}
\exp\bigg{[}-i\alpha' (U^{00}-U^{0i_{0}}V)^{-1}{(P^{0})}^{2}
\nonumber\\
&~&-2i\alpha'\sum_{\bar{\beta}}\bigg{(}
\frac{\sum_{\bar{\alpha}}[(1-\frac{1}{2}
\delta_{\bar{\alpha}\bar{\beta}})U^{\bar{\beta}
\bar{\alpha}}P^{\bar{\alpha}}]}{\sum_{\bar{\gamma}}
(U^{\bar{\beta}\bar{\gamma}}U^{\bar{\beta}
\bar{\gamma}})}P^{\bar{\beta}}\bigg{)}\bigg{]}
\nonumber\\
&~&\times\delta(x^{i_{0}}-Vx^{0}-y^{i_{0}}) \prod_{j\neq
i_{0}}\delta(x^{j}-y^{j})
\nonumber\\
&~&\times\prod_{\alpha}|p_{L}^{\alpha}=p_{R}^{\alpha}\rangle
\prod_{j\neq i_{0}}|p_{L}^{j}=p_{R}^{j}=0
\rangle|p_{L}^{i_{0}}=p_{R}^{i_{0}}=\frac{1}{2}Vp^{0}\rangle
\bigg{\}},
\eea
where $P^{\alpha}=p^{\alpha}-Vp^{i_{0}}{\delta^{\alpha}}_{0}$ and
$\bar{\alpha}, \bar{\beta}$ and $\bar{\gamma}$ take their values
from the spatial directions of the brane. The index $j$ indicates the
directions perpendicular to the brane except $i_{0}$. Since the tachyon field
prevents the closed string from wrapping around the compact directions,
left and right-components of the momentum are equal. This implies that
closed string has zero winding numbers, and its momentum components are
not discrete. Therefore, in this state there are integrals over them.
\section{Interaction of the branes}
\subsection{The amplitude}

For calculating the interaction amplitude we need the total
boundary state. What was acquired in (20) and (26) is the matter
part of it. We should also take into account the boundary state associated
with the conformal ghosts. This is due to the fact that we are
working in the covariant formalism. So the total boundary state
which will be used to calculate the amplitude, is
\bea
|B,\tau=0\rangle=|B_{\rm osc},\tau=
0\rangle{|B_{x},\tau=0\rangle}^{(0)}|B_{\rm gh},\tau=0\rangle,
\eea
where the ghost part is
\bea
|B_{\rm gh},\tau=0\rangle=\exp\bigg{[}\sum^{\infty}_{m=1}
(c_{-m}\widetilde{b}_{-m}-b_{-m}\widetilde{c}_{-m})\bigg{]}\frac{c_{0}
+\widetilde{c_{0}}}{2}|q=1\rangle|\widetilde{q}=1\rangle.
\eea

Now we proceed to calculate the interaction amplitude between the
D$p_{1}$-brane and D$p_{2}$-brane through closed string
exchanging between them. For this purpose we need closed string propagator
which is given by $D=2\alpha'\int_{0}^{\infty}dt e^{-tH} $ where
$H$ is the closed string Hamiltonian. Overlap of two boundary states,
corresponding to the branes, via this propagator defines
the interaction amplitude, i.e. ${\cal A}=\langle
B_{1}|D|B_{2}\rangle $. Before calculating the interaction
amplitude there are some conventions for indices. The set
$\{\bar{i}\}$ shows directions perpendicular to both branes except
$i_{0}$, $\{\bar{u}\}$ is for the directions along both branes
except $0$, $\{\alpha'_{1}\}$ is used for the directions along the
D$p_{1}$-brane and perpendicular to the D$p_{2}$-brane, and
$\{\alpha'_{2}\}$ indicates the directions along the D$p_{2}$-brane
and perpendicular to the D$p_{1}$-brane. Therefore, we have the
relations
\bea
&~&\{\alpha_{1}\}=\{\bar{u}\}\cup\{\alpha'_{1}\}\cup\{0\},
\nonumber\\
&~&\{\alpha_{2}\}=\{\bar{u}\}\cup\{\alpha'_{2}\}\cup\{0\},
\nonumber\\
&~& \{i_{1}\}=\{\bar{i}\}\cup\{\alpha'_{2}\}\cup\{i_{0}\},
\nonumber\\
&~& \{i_{2}\}=\{\bar{i}\}\cup\{\alpha'_{1}\}\cup\{i_{0}\},
\nonumber\\
&~& \{\mu\}=\{\alpha_1\}\cup\{i_1\}=\{\alpha_2\}\cup\{i_2\}.
\eea
Since the position of the branes are specified
by the running indices $\{\bar{\alpha}_{1}\}$ and
$\{\bar{\alpha}_{2}\}$, the D$p_{1}$ and D$p_{2}$-brane can be
parallel or perpendicular to each other.

After a long calculation the following interaction amplitude is acquired
\bea
{\cal {A}} =&~&
\frac{\alpha'\;V_{\overline{u}}}{4(2\pi)^{d_{\overline{i}}}}\;
\frac{T_{p_{1}}T_{p_{2}}}{|V_{1}-V_{2}|}\prod^{\infty}_{m=1}
\bigg{(}\det[M_{(m)1}M_{(m)2}]\bigg{)}^{-1}
\nonumber\\
&~& \times {\int_{0}}^{\infty}dt \;
\bigg{\{}e^{(d-2)t/6}\prod^{\infty}_{m=1}\bigg{(}[\det(1-{\cal{S}}_{(m)1}
{\cal{S}}_{(m)2}^{T}e^{-4mt})]^{-1}(1-e^{-4mt})^{2}\bigg{)}
\nonumber\\
&~&\times
\bigg{(}\sqrt{\frac{\pi}{\alpha't}}\bigg{)}^{d_{\bar{i}_{n}}}
\exp \bigg{[}-\frac{1}{4\alpha't}\sum_{\bar{i}_{n}}({y_{1}}
^{\bar{i}_{n}}-{y_{2}}^{\bar{i}_{n}})^{2}\bigg{]}
\prod_{\bar{i}_{c}}\Theta_{3}\bigg{(}
\frac{y_{1}^{\bar{i}_{c}}-y_{2}^{\bar{i}_{c}}}{2\pi
R_{\bar{i}_{c}}}\mid\frac{i\alpha't}{\pi
(R_{\bar{i}_{c}})^{2}}\bigg{)}
\nonumber\\
&~&\times\int^{+\infty}_{-\infty}\prod_{\alpha'_1}
{dp_{1}}^{\alpha_{1}}\prod_{\alpha'_2}{dp_{2}}
^{\alpha_{2}}\bigg{[}\exp \bigg{(}-\alpha't({\textit{f}}^
{\;(+)}{\textit{f}}^
{\;(-)}+{p_{1}}^{\alpha'_{1}}{p_{1}}^{\alpha'_{1}}+
{p_{2}}^{\alpha'_{2}}{p_{2}}^{\alpha'_{2}}
+{p_{1}}^{\bar{u}}{p_{2}}^{\bar{u}})\bigg{)}
\nonumber\\
&~&\times\exp\bigg{(}i\Phi(12){y_{2}}^{i_{0}}-
i\Phi(21){y_{1}}^{i_{0}}
+2i{y_{1}}^{{\alpha'}_{2}}{p_{2}}^{{\alpha'}_{2}}-2i{y_{2}}^
{{\alpha'}_{1}}{p_{1}}^{{\alpha'}_{1}}\bigg{)}
\nonumber\\
&~&\times\exp\bigg{(}{\pi^{0}}^{T}Q \pi^{0}+E^{T}\pi^{0}
+{\pi_{1}}^{T} G_{1} \pi_{1}+{\pi_{2}}^{T} G_{2}
\pi_{2}\bigg{)}\bigg{]} \bigg{\}},
\eea
where $V_{\overline{u}}$ is the common worldvolume.
$\bar{i}_{c}$ and $\bar{i}_{n}$ indicate the compact and
non-compact parts of $X^{{\bar i}}$, with
$d_{{\bar i}_n}={\rm dim}\{X^{{\bar i}_n}\}$ and
$d_{{\bar i}}={\rm dim}\{X^{{\bar i}}\}$.
The directions $X^{{\bar i}_c}$ and $X^{{\bar i}_n}$ are perpendicular
to both branes. In addition, $R_{\bar{i}_{c}}$ is the radius of
compactification and $d$ is the spacetime dimension. We also defined
$\pi^{0}=\left( \begin{array}{c}
p_{1}^{0} \\
p_{2}^{0}
\end{array} \right)$, $\pi_{1}=\left( \begin{array}{c}
p_{1}^{\alpha'_{1}} \\
p_{1}^{\bar{u}}
\end{array} \right) $ and $\pi_{2}=
\left( \begin{array}{c} p_{2}^{\alpha'_{2}} \\
p_{2}^{\bar{u}}
\end{array} \right) $ in which $\pi_1$ and $\pi_2$
have $(d_{\alpha'_{1}}+d_{\bar{u}})$ and
$(d_{\alpha'_{2}}+d_{\bar{u}})$ elements, respectively. The
notations $\Phi(12)$ and ${\textit{f}}^{\;(+)}$ are defined by
\bea
\Phi(12)=\frac{1}{V_{2}-V_{1}}[(1+V_{1}V_{2})p_{2}^
{0}-(1+V_{1}^{2})p_{1}^{0}],
\eea
\bea
{\textit{f}}^{\;(+)}=\frac{1}{|V_{2}-V_{1}|}[(1+V_{1})
(1+V_{2}^{2})p_{2}^{0}-(1+V_{2})(1+V_{1}^{2})p_{1}^{0}].
\eea
By exchanging $2\leftrightarrow1$ in (31) we receive $\Phi(21)$, and
by changing $V_{1}\longrightarrow-V_{1}$ and
$V_{2}\longrightarrow-V_{2}$ in (32), ${\textit{f}}^{\;(-)}$ can be
obtained. The matrices $Q$, $G_{1}$ and $G_{2}$ and the doublet
$E=\left( \begin{array}{c}
E_1\\
E_2
\end{array} \right)$,
are defined through their elements as in the following
\bea
\left\{
\begin{array}{rcl} &~& Q_{11}=
\frac{\alpha't}{(V_{2}-V_{1})^{2}}(1+{V_{1}}^{2})(1-{V_{2}}^{2})+
2i\alpha'(U^{00}_{1}
-U^{0,i_{0}}_{1}V_{1})^{-1}(1-{V_{2}}^{2})^{2},\\
&~& Q_{22}=\frac{\alpha't}{(V_{2}-V_{1})^{2}}(1+{V_{2}}^{2})
(1-{V_{1}}^{2})
-2i\alpha'(U^{00}_{2}-U^{0,i_{0}}_{2}V_{2})^{-1}(1-{V_{1}}^{2})^{2},\\
&~&
Q_{12}=Q_{21}=\frac{\alpha't}{(V_{2}-V_{1})^{2}}(1+{V_{1}}^{2})
(1+{V_{2}}^{2})(1-V_{1}V_{2}),
\end{array}\right.
\eea
\bea
\left\{ \begin{array}{rcl} &~&
E_{1}=\frac{i}{V_{2}-V_{1}}[{y_{2}}^{i_{0}}(1+{V_{1}}^{2})^{2}
-{y_{1}}^{i_{0}}(1+V_{1}V_{2})],
\\ &~& E_{2}=\frac{i}{V_{2}-V_{1}}[{y_{1}}^{i_{0}}
(1+{V_{2}}^{2})^{2} -{y_{2}}^{i_{0}}(1+V_{1}V_{2})],
\end{array}\right.
\eea
\bea \left\{ \begin{array}{rcl} &~&
G^{\alpha'_{1}\beta'_{1}}_{1}
=4i\alpha'\frac{(1-\frac{1}{2}\delta_{\alpha'_{1}
\beta'_{1}})U^{\alpha'_{1}\beta'_{1}}_{1}}{\sum_{\bar{\gamma}_{1}}
(U^{\bar{\gamma}_{1}\beta'_{1}}_{1}U^{\bar{\gamma}_{1}
\beta'_{1}}_{1})}
-\alpha't\delta^{\alpha'_{1}\beta'_{1}}, \\
&~& G^{\alpha'_{1}\bar{u}}_{1}=4i\alpha'
\frac{U^{\alpha'_{1}\bar{u}}_{1}}{\sum_{\bar{\gamma}_{1}}
(U^{\bar{\gamma}_{1}
\bar{u}}_{1}U^{\bar{\gamma}_{1}\bar{u}}_{1})}, \\
&~& G^{\bar{u}\alpha'_{1}}_{1}=4i\alpha'\frac{U^{\bar{u}
\alpha'_{1}}_{1}}{\sum_{\bar{\gamma}_{1}}(U^{\bar{\gamma}_{1}
\alpha'_{1}}_{1}U^{\bar{\gamma}_{1}\alpha'_{1}}_{1})},\\
&~& G^{\bar{u}\bar{v}}_{1}=4i\alpha'\frac{(1-\frac{1}{2}\delta_{\bar{u}
\bar{v}})U^{\bar{u}\bar{v}}_{1}}{\sum_{\bar{\gamma}_{1}}(U^{\bar{v}
\bar{\gamma}_{1}}_{1}U^{\bar{v}\bar{\gamma}_{1}}_{1})}
-\frac{1}{2}\alpha't\delta^{\bar{u}\bar{v}}.
\end{array}\right.
\eea
With the exchange $1\longleftrightarrow2$ and $i\rightarrow-i$ in the
elements of $G_{1}$ we receive the elements of $G_{2}$.
For parallel D-branes with the same dimension those terms which
contain $\alpha'_1$ and $\alpha'_2$ disappear. The effects of
compactification are in the product of the $\Theta_3$-functions.
Therefore, the amplitude in the non-compact spacetime can be obtained
as follows: remove the $\Theta_3$-functions, change
${\bar i_n} \rightarrow {\bar i}$ and $d_{\bar i_n} \rightarrow d_{\bar i}$.

We can make the amplitude simpler by performing the integration over
momenta. After introducing the regularization (25) it finds the feature
\bea
{\cal {A}} =&~&
\frac{\alpha'\;V_{\overline{u}}}{4(2\pi)^{d_{\overline{i}}}}\;
\frac{T_{p_{1}}T_{p_{2}}}{|V_{1}-V_{2}|}
\sqrt{\det(\Omega_1\Omega_2)}\;
\det\bigg{[}\Gamma \bigg{(}\frac{U_1}{2i\Omega_1}+1\bigg{)}
\Gamma \bigg{(}\frac{U_2}{2i\Omega_2}+1\bigg{)}\bigg{]}
\nonumber\\
&~& \times {\int_{0}}^{\infty}dt \;
\bigg{\{}e^{(d-2)t/6}{\prod^{\infty}_{m=1}}\bigg{(}[\det(1-{\cal{S}}_{(m)1}
{\cal{S}}_{(m)2}^{T}e^{-4mt})]^{-1}(1-e^{-4mt})^{2}\bigg{)}
\nonumber\\
&~&\times
\bigg{(}\sqrt{\frac{\pi}{\alpha't}}\bigg{)}^{d_{\bar{i}_{n}}}
\exp \bigg{[}-\frac{1}{4\alpha't}\sum_{\bar{i}_{n}}({y_{1}}^{\bar{i}_{n}}-
{y_{2}}^{\bar{i}_{n}})^{2}\bigg{]}
\prod_{\bar{i}_{c}}\Theta_{3}\bigg{(}\frac{y_{1}^{\bar{i}_{c}}-
y_{2}^{\bar{i}_{c}}}{2\pi R_{\bar{i}_{c}}}\mid\frac{i\alpha't}{\pi
(R_{\bar{i}_{c}})^{2}}\bigg{)}
\nonumber\\
&~&\times\frac{1}{\sqrt{\det Q \;\det G_{1}\;\det G_{2}}}\nonumber\\
&~&\times\exp\bigg{[}-\frac{1}{4}\bigg{(}E^{T}Q^{-1}E+
\sum_{\alpha'_{1},\beta'_{1}}
(y_{2}^{\alpha'_{1}}y_{2}^{\beta'_{1}}(G_{1}^{-1})_
{\alpha'_{1}\beta'_{1}}) +\sum_{\alpha'_{2},\beta'_{2}}
(y_{1}^{\alpha'_{2}}y_{1}^{\beta'_{2}}(G_{2}^{-1})_
{\alpha'_{2}\beta'_{2}}) \bigg{)}\bigg{]}\bigg{\}}. \eea
The
tachyon, Kalb-Ramond and gauge fields are collected in the matrices
$\Omega_1$, $\Omega_2$, ${\cal{S}}_1$, ${\cal{S}}_2$, $Q$, $G_1$,
$G_2$, and the doublet $E$. The amplitude is symmetric under the
exchange of $1\longleftrightarrow2$, as expected.

The constant factors behind the integral show the strength of the
interaction. The second line in (36) reflects the portion of
oscillators and conformal ghosts in interaction. The exponential
factor in the third line is a damping factor with respect to the
distance of the branes. If all directions $\{X^{\bar i}\}$ are
compact then $d_{{\bar i}_n}=0$ and this exponential factor
disappears. Similarly, if they are non-compact then the
$\Theta_3$-factor will be eliminated.
\subsection{Behavior of the interaction amplitude for large distances}

In any interaction theory one should verify the large distances
behavior of the amplitude. This gives long-range force of the theory.
In our case it is related to the contribution of the closed string tachyon
and massless states to the interaction. Now our aim is to
verify this statement for our system which contains a special
tachyon field. In other words, we intend to study the effect of
the background fields on the interaction amplitude after long
times. For this purpose we should perform the limit,
$\lim_{t\rightarrow\infty}{\cal {A}}$. Since the matrices $Q$,
$G_1$ and $G_2$ are functions of time, for $d=26$ there
is the following limit
\bea
&~& \lim_{t\rightarrow\infty}\bigg{\{}e^{4t}{\prod^{\infty}_{m=1}}
\bigg{(}[\det(1-{\cal{S}}_{(m)1}{\cal{S}}_{(m)2}^{T}e^{-4mt})]
^{-1}(1-e^{-4mt})^{2}\bigg{)}
\nonumber\\
&~& \times\frac{1}{\sqrt{\det Q \;\det G_{1}\;\det
G_{2}}}\exp\bigg{(}-\frac{1}{4}E^{T} Q^{-1}E
\bigg{)}\bigg{\}}
\nonumber\\
&~&=\frac{i2^{d_{\bar{u}}+1/2}\;(-1)^{(p_{1}+p_{2})/2}}
{\alpha'^{(p_{1}+p_{2})/2}}\;
\frac{|V_{1}-V_{2}|}{(1+{V_{1}}^{2})(1+{V_{2}}^{2})}
\lim_{t\rightarrow\infty}
\bigg{\{}\frac{e^{4t}}{t^{1+(p_1+p_2)/2}}+
\frac{{\rm Tr}{({\cal{S}}_{(1)1}{\cal{S}}_{(1)2}^{T})-2}}
{t^{1+(p_1+p_2)/2}}\bigg{\}}.
\eea
Substituting the limit (37) into the amplitude (36), the massless states and
tachyon contributions to the interaction amplitude for $d=26$ become
\bea
{\cal {A}}_0= &~& \frac{V_{\overline{u}}\;T_{p_1}T_{p_2}}
{4(2\pi)^{d_{\bar{i}}}}\; \sqrt{\det(\Omega_1\Omega_2)}\;
\det\bigg{[}\Gamma \bigg{(}\frac{U_1}{2i\Omega_1}+1\bigg{)}
\Gamma \bigg{(}\frac{U_2}{2i\Omega_2}+1\bigg{)}\bigg{]}
\nonumber\\
&~&\times \frac{i\;(-1)^{(p_{1}+p_{2})/2}\;2^{d_{\bar{u}}+1/2}}
{\alpha'^{(p_{1}+p_{2})/2}}\;
\frac{1}{(1+{V_{1}}^{2})(1+{V_{2}}^{2})}\;
\nonumber\\
&~& \times \int^\infty dt\;\bigg{\{}
\bigg{(}\sqrt{\frac{\pi}{\alpha't}}\bigg{)}^{d_{\bar{i}_{n}}}
\exp \bigg{[}-\frac{1}{4\alpha't}\sum_{\bar{i}_{n}}({y_{1}}^{\bar{i}_{n}}-
{y_{2}}^{\bar{i}_{n}})^{2}\bigg{]}
\prod_{\bar{i}_{c}}
\Theta_{3}\bigg{(}\frac{y_{1}^{\bar{i}_{c}}-y_{2}^{\bar{i}_{c}}}{2\pi
R_{\bar{i}_{c}}}\mid\frac{i\alpha't}{\pi
(R_{\bar{i}_{c}})^{2}}\bigg{)}
\nonumber\\
&~& \times \lim_{t\rightarrow\infty}
\bigg{[}\frac{e^{4t}}{t^{1+(p_1+p_2)/2}}+
\frac{{\rm Tr}{({\cal{S}}_{(1)1}{\cal{S}}_{(1)2}^{T})-2}}{t^{1+
(p_1+p_2)/2}}\bigg{]}\bigg{\}}.
\eea
The divergent part in the last bracket corresponds to the tachyonic closed
string state. It differs from the same part in the papers by the
coefficient $1/{t^{1+(p_1+p_2)/2}}$ which slows down
this divergence. The other term of the last bracket
is related to the contribution of
the massless fields which goes to zero fast in the limit of large
distances. It is notable that this damping factor just depends on
the two D-branes dimensions not their relative configuration. So
in the presence of the tachyon field the behavior of the interaction
amplitude has changed in such a manner that in large distances the
contribution of the graviton, dilaton and Kalb-Ramond fields
disappears. This effect may be understood as follows.

According to the first equation of (17) the energy $p^{0}$ defines
a linear potential, acting on the closed string. This potential
completely originates from the background tachyon. This potential
slows down the closed string motion, which happens for all closed
string states, including the massless states. Therefore, the
exchanged closed strings will cease and hence there is no long
range force.

Vanishing of the D-branes interaction after long enough time also
may be interpreted as rolling of the tachyon field in this limit
\cite{19}. Presence of the open string tachyon field implicates the
instability of the D-brane. That is, in long time limit the tachyon
rolls down to its minimum potential and hence the D-brane decays to
closed string states and finally disappears. Thus, after long time
there are no D-branes to interact with each other.
\section{Conclusions}

We obtained the boundary state of a closed string, emitted
(absorbed) from (by) a moving brane in the presence of the
background fields $B_{\mu\nu}$, tachyon and internal $U(1)$ gauge
field. By partially compactifying the spacetime on tori,
the formalism was applied to both compact and
non-compact spacetime. We observed that the closed string
can not wrap around the compact directions of the spacetime which are
perpendicular to the brane's worldvolume. In addition, for a special
tachyon matrix, the tachyon prevents the closed string from winding
around the compact directions parallel to the brane's volume.

The interaction amplitude of two D-branes with arbitrary dimensions
was calculated. The D-branes are parallel or perpendicular to each
other. Due to the tachyon field, the interaction strength
between the branes depends on all mode numbers of the exchanged
closed string. The background fields, specially the tachyon
field, affect the interaction. In other words, the background fields
define an effective tension for each D-brane. In the corresponding
potential, related to the interaction amplitude, the product of these
effective tensions define a coupling constant. The value of this
coupling constant indicates the strength of the interaction. Therefore,
by adjusting the parameters $V_1$, $V_2$, $\{U^{(1)}_{\mu\nu},U^{(2)}_{\mu\nu}\}$,
$\{{\cal{F}}^{(1)}_{\mu\nu}, {\cal{F}}^{(2)}_{\mu\nu}\}$ and
$\{R^\mu |\mu \neq 0 \}$ we can control this strength.

In the large distance of the
branes, the contribution of the massless states (i.e. graviton, dilaton
and Kalb-Ramond fields) goes to zero and the
divergence part related to the tachyonic closed string state
considerably slows down. This can be understood by the decelerating
potential which acts on the exchanging closed strings, and also by
rolling of the tachyon which leads to the instability of the D-branes
and hence decaying of them.

Although in this article we are dealing only with bosonic
string, it is worth noting that similar consideration,
concerning the effects of compactification, works for
the superstring case since these effects are independent
of the fermions. However, adding fermionic degrees of freedom to
the present formalism is in progress.

\end{document}